\documentclass{article}

\usepackage{arxiv}
\usepackage{amsmath, amssymb}
\usepackage{tabularx}
\usepackage{listings}
\usepackage[utf8]{inputenc} 
\usepackage[T1]{fontenc}    
\usepackage[hidelinks]{hyperref}       
\usepackage{url}            
\usepackage{booktabs}       
\usepackage{amsfonts}       
\usepackage{nicefrac}       
\usepackage{microtype}      
\usepackage{lipsum}		
\usepackage{graphicx}
\usepackage{natbib}
\usepackage{doi}
\usepackage{bm}
\usepackage{amsmath}
\usepackage{xcolor}
\newcommand{\tr}{^\intercal}

\title{A Nonparametric Bayesian Model to Adjust for Monitoring Bias with an Application to Identifying Environments Stressed by Climate Change}

\author{
    \href{https://orcid.org/0000-0003-3772-8030}{\includegraphics[scale=0.06]{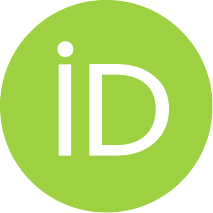}\hspace{1mm}\textnormal{Jonathan Auerbach}} \\
    Department of Statistics \\
    George Mason University \\
    Fairfax, VA 22030 \\
    \texttt{jauerba@gmu.edu} \\
    \and
    \href{https://orcid.org/0000-0001-9592-625X}{\includegraphics[scale=0.06]{orcid.pdf}\hspace{1mm}Theresa M. Crimmins} \\ 
    USA National Phenology Network \\
    School of Natural Resources and the Environment \\
    University of Arizona \\
    Tucson, AZ 85721 \\
    \texttt{theresam@u.arizona.edu} \\
    \and 
    \href{https://orcid.org/0000-0002-3776-4819}{\includegraphics[scale=0.06]{orcid.pdf}\hspace{1mm}David Kepplinger} \\
    Department of Statistics \\
    George Mason University \\
    Fairfax, VA 22030 \\
    \texttt{dkepplin@gmu.edu} \\
    \and 
    \href{https://orcid.org/0009-0005-0275-1822}{\includegraphics[scale=0.06]{orcid.pdf}\hspace{1mm}Ruishan Lin} \\
    Department of Statistics \\
    George Mason University \\
    Fairfax, VA 22030 \\
    \texttt{rlin8@gmu.edu} \\
    \and 
    \href{https://orcid.org/0000-0001-7653-893X}{\includegraphics[scale=0.06]{orcid.pdf}\hspace{1mm}E.M. Wolkovich} \\
    Forest and Conservation Sciences \\
    University of British Columbia \\
    Vancouver, BC, Canada, V6T 1Z4 \\
    \texttt{e.wolkovich@ubc.ca} \\
}

\date{}

\renewcommand{\undertitle}{}
\renewcommand{\shorttitle}{A Nonparametric Bayesian Model to Adjust for Monitoring Bias}
\renewcommand{\headeright}{Auerbach et al.}

\begin{document}
\maketitle

\begin{abstract}
    We propose a new method to adjust for the bias that occurs when an individual monitors a location and reports the status of an event. For example, a monitor may visit a plant each week and report whether the plant is in flower or not. The goal is to estimate the time the event occurred at that location. The problem is that popular estimators often incur bias both because the event may not coincide with the arrival of the monitor and because the monitor may report the status in error. To correct for this bias, we propose a nonparametric Bayesian model that uses monotonic splines to estimate the event time. We first demonstrate the problem and our proposed solution using simulated data. We then apply our method to a real-world example from phenology in which lilac are monitored by citizen scientists in the northeastern United States, and the timing of the flowering is used to study anthropogenic warming. Our analysis suggests that common methods fail to account for monitoring bias and underestimate the peak bloom date of the lilac by 48 days on average. In addition, after adjusting for monitoring bias, several locations had anomalously late bloom dates that did not appear anomalous before adjustment. Our findings underscore the importance of accounting for monitoring bias in event-time estimation. By applying our nonparametric Bayesian model with monotonic splines, we provide a more accurate approach to estimating bloom dates, revealing previously undetected anomalies and improving the reliability of citizen science data for environmental monitoring.
\end{abstract}


\section{Introduction}
We consider the problem of estimating the time of an event which cannot be observed directly. Instead, individual observers make multiple visits to the location where the event is to take place and record whether the event has already occurred or not at each time of their arrival. The records are then used to estimate the event time. 

We are motivated by a real-world problem in phenology, the field which studies the timing of seasonal life cycle events in plants and animals \citep{Forrest2010}. For recent applications, see \citep{Dennis2024} and \citep{Schwob2023dynamic}. Citizen scientists monitor the status of flowering (flowers open or not open) of lilac in the northeastern United States, and anomalously early or late bloom dates are used as evidence of environmental stress caused by climate change. Typically, bloom dates are considered early or late relative to historic records before major anthropogenic warming began in 1980 \citep{stoker2013climatechange}. The problem with these comparisons is that changes in how citizen scientists monitor a site can make the site appear more anomalous than it actually is.

We focus on the peak bloom date, which is the first date when more than 50\% of the flowers have bloomed for a plant in a location. We consider two sources of variation. The first is that the citizen scientists may not arrive on the day that the lilac bloom. The second is that the citizen scientists may report the flowering status of the lilac in error. We show that when these sources of variation are not properly accounted for, substantial bias can result.

We present our work in five sections. In Section \ref{sec:background}, we review how the field of phenology relies on "status monitoring" to identify environments stressed by climate change. The purpose of this section is to further motivate the monitoring bias problem and our proposed solution. However, the problem is not unique to phenology. Correcting for monitoring bias is a general problem in the agricultural, biological, and environmental sciences. In Section \ref{sec:demonstration}, we state our assumptions and provide a demonstration of monitoring bias. In particular, we show that the size of the bias depends on the frequency with which the monitor visits the site. In Section \ref{sec:model}, we describe our solution. We propose a nonparametric Bayesian model that jointly models citizen science data across multiple sites.
 
In Section \ref{sec:application}, we apply the proposed model to a real-world monitoring problem in which citizen scientists monitored the status of common lilacs (\emph{Syringa vulgaris}) in 2022 at 25 sites throughout the northeast U.S. and reported the observed status of open flowers to the USA National Phenology Network. We find that failing to correct for monitoring bias underestimates the peak bloom date of the lilac by 48 days on average. We also find that, after adjusting for monitoring bias, 2 of the 25 sites had anomalously late bloom dates. These sites did not look atypical before adjusting for monitoring bias, and the proposed model allows us to quantify how certain we are that these sites are in fact anomalous. We conclude in Section \ref{sec:discussion} with a discussion.

\section{Background}\label{sec:background}

Climate change impacts environments in complicated ways that are difficult to observe directly. Scientists have increasingly turned to data collected by citizen scientists to better understand these impacts. A key source of data comes from the volunteers who monitor the seasonal activity of plants and animals around the world \citep{Battle2022, Menzel2020}. A major strength of citizen science data is the quantity of observations that can be amassed. Citizen science data provide considerably more information than what can be collected by paid scientists alone. 

In the United States, seasonal activity in plants and animals is documented by volunteers and professional scientists at thousands of locations through the USA National Phenology Network \emph{Nature's Notebook} platform \citep{rosemartin2014organizing}. Since the program's launch in 2009, over 40 million records have been contributed, and these data have been used in hundreds of scientific analyses and to support decision-making in a wide range of applications \citep{crimmins2022science}.

As statistical methods for analyzing phenological data become more critical to science and policy, understanding and correcting their biases is increasingly important. We show that current methods used to determine the timing of seasonal onsets are sensitive to a bias that we refer to as monitoring bias. The bias occurs because observers use ``status monitoring'' to document what they see: each time they visit a site, they report whether they see plants or animals expressing a series of life cycle stages called phenophases \citep{denny2014standardized}. For example, a monitor might record whether they observe a particular plant has open flowers on the date they make the observation. They do not record the date that the plant bloomed. Common estimates of the bloom date, such as the first day the monitor observes a bloom, are biased both because the monitor rarely arrives on the actual bloom date but also because the monitor may report the bloom date in error. Indeed, it is not uncommon for a monitor to report the status of a plant as bloomed on one visit only to reverse the status from bloomed to not bloomed on a subsequent visit. This may suggest that either the first or second report was in error, although it is also possible for the plant to reach flowering status only for an environmental changes, such as cold weather, to revert the status back to unflowered.

Our key assumption is that each monitor is unbiased in the sense that, were a monitor to evaluate a large number of plants at a site, the monitor would correctly report the status on average. Yet even under this unbiased assumption, we find that methods commonly used to estimate the bloom date in phenology can be severely biased. Moreover, the size of the bias depends on the frequency with which a site is monitored so that differences in the timing or frequency with which the monitor visits the site can make the environment appear more or less anomalous than it actually is.

We show that the bias may be corrected by fitting a model to the data and estimating the typical bloom date from the fitted model, provided the model is correctly specified. However, a misspecified model can introduce more bias than the naïve approach of simply using the first day the monitor reported a bloom. It is for this reason that we propose a nonparametric Bayesian model. Our model can be seen as a compromise between the parametric approach --- which estimates the underlying distribution of bloom dates to determine the typical bloom date --- and the naïve approach --- which does not assume the underlying distribution is known.

\section{Assumptions and demonstration}\label{sec:demonstration}

We begin with a demonstration of monitoring bias. The purpose is to motivate the main assumptions, the parameter of interest, and the proposed approach. We therefore limit our attention to the case in which one monitor repeatedly visits a single site. In Subsection \ref{subsec:assumptions}, we state our primary assumptions. In Subsection \ref{subsec:demonstration}, we provide the demonstration. In Section \ref{sec:model}, we generalize to the case where each site is visited by multiple monitors.

\subsection{Assumptions} \label{subsec:assumptions}

Let $Y_i$ denote a binary random variable, indicating whether a monitor reported a plant to have bloomed or not upon the $i$th visit to a site, $i = 1, \ldots, n$. We make two assumptions. The first assumption (A1) is that the $\{ Y_i \}_{i=1}^n$ are independent Bernoulli random variables, i.e., $Y_i \sim \text{Bernoulli}(p_i)$. The second assumption (A2) is that a monitor is unbiased in the sense that, were the monitor to hypothetically evaluate a large number of plants at the site, the status reported by the monitor would be correct on average. 

The first assumption implies that each status report $Y_i$ is based solely on information available at the $i$th visit and does not depend on the report from previous visits. Consequently, it is possible for the status reports to be inconsistent: for example, the monitor might report that a plant has bloomed on the first visit ($y_1 = 1$) but then report that it has not bloomed on the second visit ($y_2 = 0$). Such contradictions are common 9when citizen scientists engage in status monitoring and indicate that either 1) the status changed, or 2) one or both reports were made in error. (Here and throughout this paper, $y_i$ denotes a realization of the random variable $Y_i$.)

Our second assumption is that, though the reports may be made in error, a monitor is correct on average. Consider a scenario where a large number of plants are present at the site. Let $F(t)$ denote the proportion of these plants that would have bloomed by day $t$. (i.e., $F(t)$ is the cumulative distribution function of bloom dates, or bloom distribution for short.) We assume $p_i = F(t_i)$, where $t_i$ denotes the day of the $i$th visit. 

The second assumption reflects the `wisdom of the crowds' phenomenon. We argue reported status is random in part because monitors use context clues, such as the status of other plants during the visit, to determine whether the definition of a bloom has been met. We assume these context clues vary by visit but do not bias the judgment of the citizen scientist towards a particular status. We believe this explains the `wisdom of the crowds' --- that while individuals may be unreliable, the crowd is often accurate --- in many cases more accurate than an expert.

\subsection{Demonstration} \label{subsec:demonstration}

\begin{figure*}[!t]
\centering
\includegraphics[width=\linewidth]{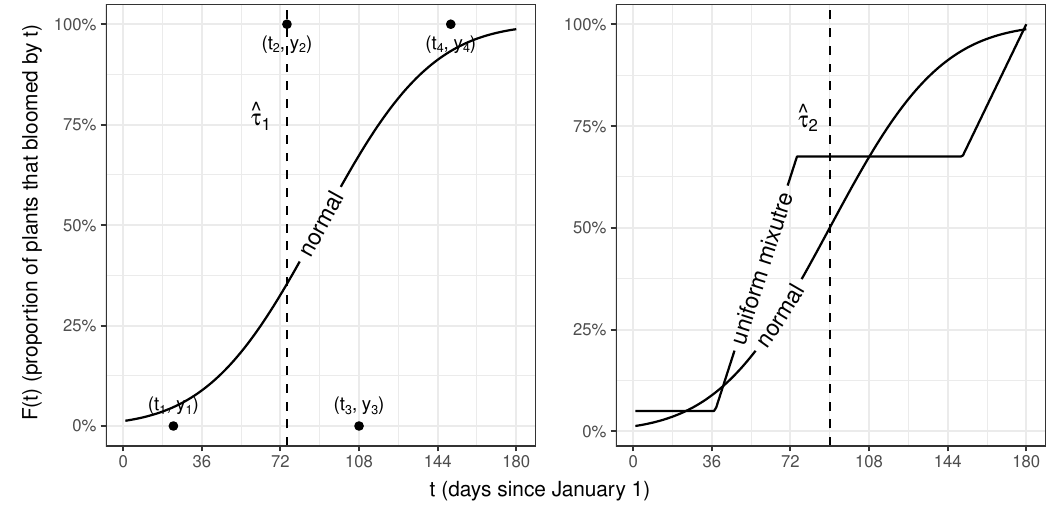}

\caption{An illustration of monitoring bias. Left panel: A monitor visits a site on days $t_i$, $i = 1, \ldots, 4$, and reports $Y_i \sim \text{Ber}(F(t_i))$ (points), where $F(t)$ (solid line) represents the unknown proportion of plants that bloomed by time $t$ (in the figure a normal distribution). Of interest is the median $\tau = \text{inf} \{ t : F(t) > .5 \}$.
The naïve estimate $\hat \tau_1$, the first time the monitor reports a bloom (dashed line), is biased in general. Right panel: Fitting a probit model to the data yields estimate $\hat \tau_2$, which has low bias provided the underlying distribution is normal. But if bloom times follow a mixture of uniform distributions, for example, $\hat \tau_2$ may have greater bias than $\hat \tau_1$.}
\label{Figure:Figure1}
\end{figure*}

Having reviewed our assumptions, we now demonstrate the problem of monitoring bias. In the left panel of Figure \ref{Figure:Figure1}, the points represent the observed data $(t_i, y_i), i = 1, \ldots , 4$ and the solid line labeled `normal' represents one possible underlying distribution function, $F(t)$, in this case a normal distribution with mean (median) 60 and standard deviation 10. We measure $t_i$ as the number of days since January 1st, where $0 \leq t_i \leq 181$.

The parameter of interest is the peak bloom date, which we define to be the median \begin{equation} \tau = \text{inf} \{ t : F(t) > .5 \} \end{equation} We compare three methods of estimating $\tau$. The first method estimates the peak bloom date by the first day the bloom is reported: \begin{equation} \hat{\tau}_1 = \text{min} \{ t_i : Y_{i} = 1 \}. \end{equation} We refer to this estimator as the naïve approach, and it is represented by the dashed line in the left panel of Figure \ref{Figure:Figure1}. The naïve approach is widely used in practice --- likely due to its simplicity --- and may appear justified on the basis that the reported status is (assumed to be) an unbiased estimate of the distribution function. Thus, the event $Y_{i} = 1$ suggests $F(t_i) > .5$, with no additional assumptions necessary about $F$. 

Yet despite its popularity, $\hat \tau_1$ is not an unbiased estimate of $\tau$ in general. The size of the bias depends on the monitoring pattern, $t_i$, and the underlying distribution, $F$. Infrequent monitoring typically overestimates the peak bloom date, while frequent monitoring underestimates it. To see the first case, consider the extreme example where all plants at a site bloom on the same day, chosen at random between 0 and 180. A monitor that visits the site the same day each week (say Fridays) will report the bloom date 3.5 days late on average, while a monitor that chooses a random day of the month (say the 15th) will report the bloom date two weeks late on average. See related discussions of ``weekend bias'' \citep{Cooper2014, Courter2013} and censoring generally \citep{Elmendorf2019}.

To see the second case, note that if a site is monitored enough times over a short interval of time, a bloom will eventually be reported, provided that $F$ is not zero over that interval. In the extreme case, we would expect one report among a million monitors to reflect the one-in-a-million plant that just so happened to bloom in the first week. This one-in-a-million report may be considered correct in the sense that a plant at the site really did bloom in the first week. However, the reported bloom date would not be representative of the typical plant in the environment being monitored. See \citet{Moussus2010} for a related discussion on the bias of first appearance dates.

The consequence of monitoring bias is that two sites may have the same distribution of bloom times, but the naïve approach may make the environments appear different if different monitoring patterns are chosen for each site.

One solution is to fit a generalized linear model, estimate $F$, and then use the estimate to calculate the peak bloom date. The generalized linear model requires that $F$ belongs to a known parametric family, such as the family of normal distributions. This is the case with the probit model, the second method we consider:
\begin{align*}
Y_i &\sim \text{Ber}(p_i) \\
p_i &= \Phi(\alpha_0 + \alpha_1 t_i),
\end{align*}
where $\Phi$ denotes the standard normal distribution. 

The parameters $\alpha_0 \text{ and } \alpha_1$ can be estimated using maximum likelihood --- provided there is not perfect separation --- and the estimated median is \begin{equation*} \hat \tau_2 = \text{inf} \{ t : \hat{F}_2(t) > .5 \} \end{equation*} where $\hat{F}_2 = \Phi(\hat \alpha_0 + \hat \alpha_1 t)$. If there is perfect separation, $F$ can be estimated by adding prior information or regularization, see \citet{gelman2008weakly}.

Compared with $\hat \tau_1$, $\hat \tau_2$ is likely to be more accurate if a normal distribution is a good approximation of $F$, the true but unknown distribution function. However, there is nothing particularly special about the normal distribution. Other link functions are also appropriate if they approximate $F$ well. It is for this reason that the last method we consider is a compromise between the naïve approach, which makes no assumption about $F$, and the generalized linear model approach, which assumes $F$ is known to belong to a particular family. Specifically, we consider the nonparametric generalized linear model
\begin{align*}
Y_i &\sim \text{Ber}(p_i) \\
p_i &= \Phi(g(t_i)), \\
\end{align*}
where $\Phi$ denotes the standard normal distribution and $g$ is an unknown smooth monotonic link function. The function $g$ can be estimated with monotonic P-splines \citep{Pya&Wood2015} using the \texttt{R} package \texttt{scam} \citep{Pya2023}, and the median can be estimated using $\hat \tau_3 = \text{inf} \{ t : \hat{F}_3(t) > .5 \}$, where $\hat{F}_3(t) = \Phi(\hat g(t))$.

We compare all three methods in two simulation studies. In the first study, we assume the true distribution function $F$ is the normal distribution with mean $90$ and standard deviation $40$ so that the probit model is correctly specified. The bloom dates at a site might follow a normal distribution if each plant at the location is subject to a large number of small and idiosyncratic differences in growing conditions.

In the second simulation study, we assume the true distribution function is a mixture of three uniform distributions, given by the distribution function

\begin{equation} \label{eq:F(t)}
    F(t) = \begin{cases}
        0     & \quad\text{if } t = 0\\
        1/20 & \quad\text{if } 0 < t \leq 37.5\\
        3(t - 34.5) / 180 & \quad\text{if } 37.5 < t \leq 74.5 \\
        2/3 & \quad\text{if } 74.5 < t \leq 150 \\
        2(t - 90) / 180 & \quad\text{if } 150 < t \leq 180 \\
        1 & \quad\text{if } t > 180
    \end{cases}.
\end{equation}

This distribution is a mixture of three uniform distributions with domain $0$, $(37.5, 74.5]$, and $(150, 180]$. Bloom dates might follow a mixture of uniform distributions if, unlike the first simulation, the plants at a location are subject to three distinct growing conditions: for example, plants in direct sunlight might bloom first, followed by plants in a partially shaded area, followed by plants in a completely shaded area.

For both studies, we ran 1,000 simulations. In each simulation, we first sampled the monitoring times $t_i$ without replacement from the integers $\{ 1, \ldots, 180 \}$, where $i = 1, \ldots, 40$. We then sampled monitor reports $Y_i \sim \text{Ber}(F(t_i))$ and estimated the median using each of the three methods outlined above. We then repeated the two studies with $i = 1, \ldots, 50$ and $i = 1, \ldots, 60$. The results are shown in Table \ref{tab1} and \ref{tab2}. The rows indicate the number of visits in the simulations: 40, 50, and 60 visits, respectively.

\begin{table*}[t]
\caption{Bias and root mean squared error (RMSE) for three different approaches when bloom dates follow a normal distribution.\label{tab1}}
\tabcolsep=0pt
\begin{tabular*}{\textwidth}{@{\extracolsep{\fill}}lcccccc@{\extracolsep{\fill}}}
\toprule%
& \multicolumn{3}{@{}c@{}}{Bias} & \multicolumn{3}{@{}c@{}}{RMSE} \\
\cline{2-4}\cline{5-7}%
n & Naïve & Probit & Splines & Naïve & Probit & Splines \\
\midrule
40  & 41.4  & 1.2 & 6.3 & 46.4 & 11.2 & 22.2\\
50  & 46.4 & 0.6  & 4.8  & 50.3 & 9.4 & 19.2\\
60  & 49.9 & 0.5 & 3.6  & 53.3 & 8.7 & 16.1\\
\hline
\end{tabular*}
\end{table*}

\begin{table*}[t]
\caption{Bias and root mean squared error (RMSE) for three different approaches when bloom dates follow a mixture of uniform distributions.\label{tab2}}
\tabcolsep=0pt
\begin{tabular*}{\textwidth}{@{\extracolsep{\fill}}lcccccc@{\extracolsep{\fill}}}
\toprule%
& \multicolumn{3}{@{}c@{}}{Bias} & \multicolumn{3}{@{}c@{}}{RMSE} \\
\cline{2-4}\cline{5-7}%
n & Naïve & Probit & Splines & Naïve & Probit & Splines \\
\midrule
40  & 21.9  & 25.7 & 4.7 & 30.1 & 30.6 & 28.0\\
50  & 27.2 & 26.3  & 3.9  & 33.7 & 29.6 & 28.2\\
60  & 30.1 & 25.7 & 3.3  & 35.7 & 28.5 & 24.8\\
\hline
\end{tabular*}
\end{table*}

In the first simulation study, we found Probit (the parametric generalized linear model) had essentially zero bias, while Splines (the nonparametric generalized linear model) had a bias of around 5 days and the Naïve (the first reported bloom date) had a bias of around 45 days. Probit also had the lowest root mean squared error: Probit was off by about 9 days, while Splines was off by about 19 days and Naïve was off by about 50 days. This suggests that the generalized linear model performs well when the underlying distribution $F$ is known.

In the second study, we found that Splines had by far the lowest bias of around 4 days, while Naïve and Probit had a bias of more than 20 days. The root mean squared errors of the three models were largely equivalent, although Probit was less accurate than Naïve when there were 40 monitors. This suggests that the generalized linear model may not perform well when the underlying distribution $F$ is not known and in fact can perform worse than the Naïve approach.

We conclude that a nonparametric generalized linear model can provide the same or better accuracy than the naïve approach currently used to summarize citizen science data while greatly reducing bias. Note that we have only modeled one monitor at one site. Bias reduction is particularly important if researchers plan to combine data from multiple monitors at multiple sites --- for example the common practice of reporting the average estimated bloom date for a country or region --- since bias may persist when the data are aggregated across space or time.

\section{A nonparametric Bayesian model for status monitoring data}
\label{sec:model}

We now describe the proposed nonparametric Bayesian model motivated in Section \ref{sec:demonstration}. We allow for multiple monitors at multiple sites, and thus we introduce a slight change in notation. 

For each site $j = 1, \dotsc, J$ and visit $i = 1, \dotsc, n_j$, let $m_{ij}$ denote the number of monitors that visited site $j$ on day $t_{ij}$. Let $Y_{ij}$ denote the number of monitors who reported that an event occurred (e.g. the number that observed a bloom). Following Section \ref{sec:demonstration}, we assume
\begin{align*}
    Y_{ij} &\sim \text{Binomial}(m_{ij}, p_{ij}), \\
    p_{ij} &= F_j(t_{ij}).
\end{align*}

We model the marginal distribution functions using monotonic B-splines \citep{Eilers&Marx1996, Wahba1983, Wegman&Wright1983, Ramsay1988, DeBoor1978, Wahba1978}
\begin{equation}\label{eqn:model-marginal-log-likelihood}
    F_j(t) = \text{logit}^{-1}(\mathbf{B}_{t}\tr \mathbf{S} \Tilde{\bm{\beta}}_j),
\end{equation}
where $\mathbf B_{t}$ is the B-spline basis evaluated at day $t$.
Monotonicity is achieved through the shape-constraint matrix $\mathbf S$ and non-negative coefficients for the $q$-dimensional spline basis, $\Tilde{\bm{\beta}}_j = [\beta_{j1}, \exp(\beta_{j2}), \dotsc, \exp(\beta_{jq})]\tr$.
The intercept term $\beta_{j1}$ is allowed to be negative.
$\mathbf{S}$ is a square lower-triangular matrix with 
\begin{align*}
    S_{kl} = \begin{cases}
        1 \quad\text{if } k \geq l\\
        0 \quad\text{if } k < l
    \end{cases},
    \quad
     1 \leq k,l \leq q,
\end{align*}
which transforms the B-spline basis to a monotone B-spline basis.

The log-likelihood for the binomial model is given by \[\ell(\bm \beta) =\ell(y_{ij} | p_{ij}) = \sum_{j=1}^J\sum_{i=1}^{n_j} \log(p_{ij}) y_{ij} + \log(1-p_{ij}) (m_{ij} - y_{ij}).\] Note that $p_{ij}$ depends on $\bm \beta$ via (\ref{eqn:model-marginal-log-likelihood}), although this dependence is not reflected in the notation.

To enforce a certain degree of smoothness of the marginal splines, we add a penalty term to the model's log-likelihood, which penalizes the magnitude of the second derivative of the fitted function. That is, we include a smoothing parameter $\lambda$ and construct the following penalized likelihood
\begin{equation*} \label{eqn:likelihood}
    \ell(\bm{\beta}) - \lambda \bm{\beta}\tr \mathbf{P} \bm{\beta} / 2,
\end{equation*}
where $\mathbf{P} = \tilde{\mathbf{P}}\tr \tilde{\mathbf{P}}$ is a penalization matrix with
$$
\tilde{P}_{kl} = \begin{cases}
  1 & \text{if } 2 \leq k = l < q \\
 -1 & \text{if } 2 \leq k = l - 1 \leq q \\
  0 & \text{o.w.}
\end{cases}.
$$

We estimate $\tilde{\bm\beta}$ and $\theta$ in a Bayesian framework. We sample from the posterior distributions and estimate the parameters using the draw that maximizes the log posterior. The specification of priors on $\tilde{\bm\beta}$ and $\theta$ as well as other computational details are provided in Subsection \ref{subsec:data}.

\section{Application} 
\label{sec:application}

We apply the model outlined in Section \ref{sec:model} to the data described in Section \ref{sec:background}. In Subsection \ref{subsec:data}, we describe how we used the statistical software Stan \citep{Carpenter2017, RstanPackage} to fit the model to data from thirty locations that were monitored by citizen scientists throughout the northeast United States in 2022. In Subsection \ref{subsec:bias}, we use the fitted model to determine which sites if any  had anomalous bloom dates. We find that after adjusting for monitoring bias, two of the 25 sites had anomalously late bloom dates. (These sites did not look atypical before adjusting for monitoring bias.) We conclude that these sites should be examined further as they may be differentially impacted by climate change.

\subsection{We model the bloom dates of lilacs monitored by citizen scientists}
\label{subsec:data}

We use our model to examine a publicly available dataset provided by the USA National Phenology Network, detailing reports from citizen scientists monitoring the status of common lilac. We limit our investigation to sites in year 2022 in the northeastern United States that reported the flowering status at least ten times, resulting in a total of 25 sites. 

The bloom date of the common lilac is important both because of its central role in monitoring the status of spring and because it is a modern descendant of perhaps the earliest instance of citizen science, dating back to when Quetelet founded an international network for “observations of the periodical phenomena” \citep{Auerbach2023, demaree2011periodical}. The lilac are also of economic and cultural significance to the northeast United States, for example, the city of Rochester, New York is home to an annual lilac festival.

To fit the model described in Section \ref{sec:model}, we estimate the marginal distributions $F_j$ using a cubic B-spline basis of dimension $q=8$. We assume the bloom dates range between day 1 and day 180, and we position knots on an equidistant grid of the form $[0, 36, 72, \dotsc, 144, 180]$. We specify a $N(0, 50)$ prior for $\Tilde\beta_1$ and a $\text{Gamma}(1.3, 0.3)$ prior for the non-negative $\Tilde\beta_{k}$, $k > 1$. The $\theta$'s fall within the range of $[1,\infty)$, and we specify the prior $1 + \text{lognormal}(0,1)$. Additionally, we fix $\lambda_j$ at the values estimated by the \texttt{R} package \texttt{scam} (see details in \citet{Pya&Wood2015, Pya2023}).

The Stan code is available in the Appendix. We run 10 independent chains for a total of 3,000 iterations each. The initial 2500 iterations serve as a warm-up phase and were discarded. The diagnostic parameters $\hat R$ are below 1.1 for all parameters, indicating the chains have been sufficiently mixed \citep{gelman2008weakly}. Other diagnostics reported by Stan also suggest the samples are suitable for analysis.

The benefit of Stan is the flexibility with which we can specify our priors, as well as the fact that we will consider somewhat functions of the parameters when assessing whether a site is anomalous in Subsection \ref{subsec:bias}. Calculating uncertainty intervals for these functions is straightforward.

\subsection{We identify two sites with anomalously late bloom dates}
\label{subsec:bias}

\begin{figure}[t]
\centering
\includegraphics[width=0.5\linewidth]{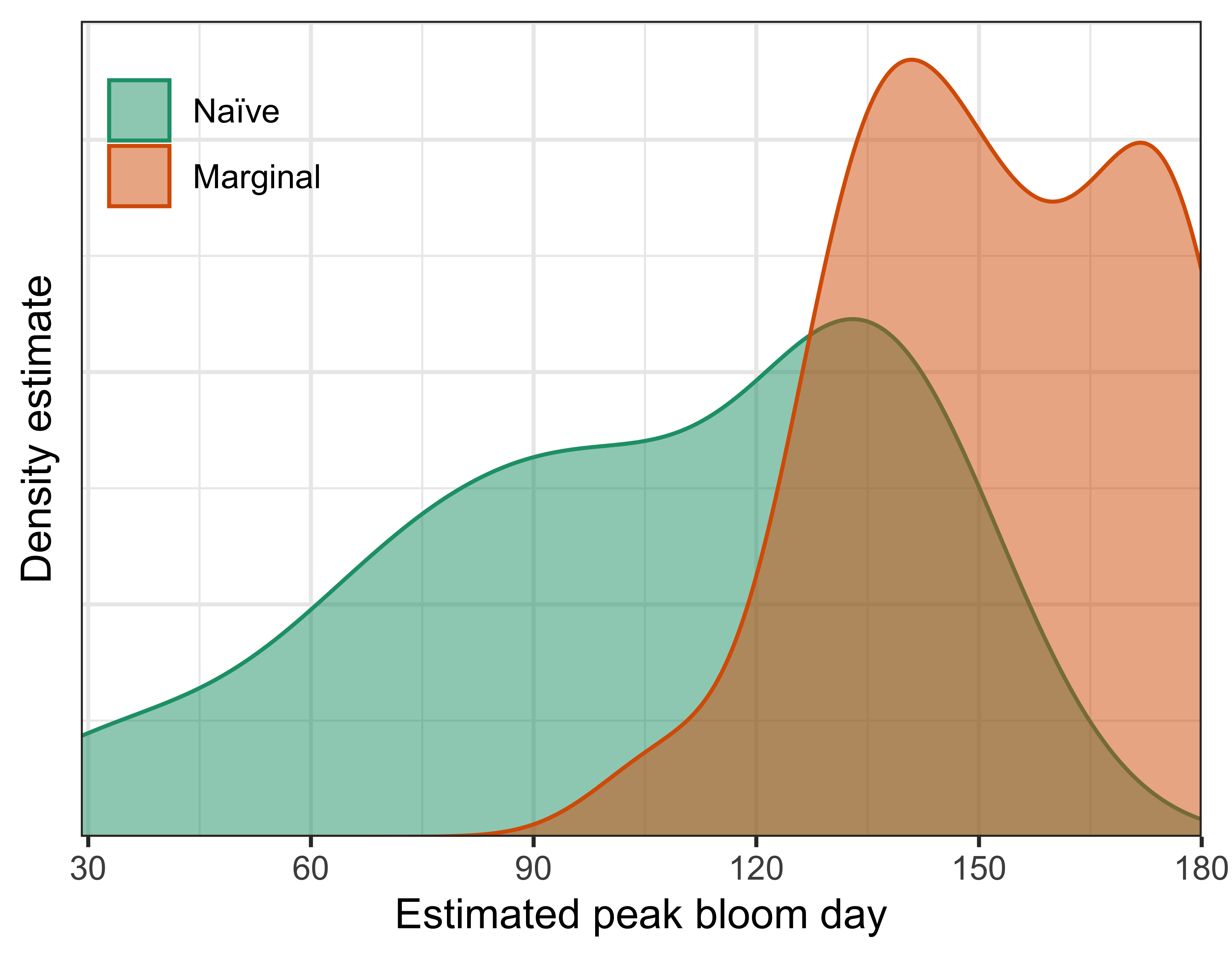}

\caption{%
Density plots of the estimated peak bloom days. The green density represents the estimated peak bloom days from the naïve method. The orange densities show estimates from the proposed approach.%
}
\label{fig:bias-joint-model}
\end{figure}

The proposed model reveals important information about the sites monitored by the citizens scientists, and this information is not captured by the naïve approach (using the first reported bloom date). The difference between the two methods is summarized in Figure~\ref{fig:bias-joint-model}. The y-axis of the figure denotes the density of the peak bloom dates estimated by the proposed approach (Proposed, orange) and the naïve approach (Naïve, green). The x-axis denotes the number of days from January 1 (day 1) until the peak bloom day.

The Naïve approach suggests most lilac bloom between late March to early May, whereas the Proposed approach suggests most lilac bloom between late April to early June. The latter is consistent with historic data about lilac bloom in the northeast, suggesting the Naïve is substantially biased. If the proposed model is correct, the Naïve approach underestimates the bloom date by 48 days on average.

If late March to early May is typical, Figure \ref{fig:bias-joint-model} shows that after adjustment, several sites had peak bloom dates in late June, which could be considered anomalously late and preliminary evidence of environmental stress. To determine the extent to which these sites are anomalous, we propose a measure based on the principal components of the covariance matrix of the bloom dates. A strength of this measure is that it does not just depend on the estimated peak bloom date, but the entire estimated distribution of bloom dates.

To construct the measure, we first estimate the covariance matrix $\hat{\bm\Sigma}$ of the bloom dates between sites. We assume the all lilac bloom by July $t = 180$, and we estimate the variance of bloom dates at site $j$ as
\begin{equation*}
    \hat{\bm\Sigma}_{jj} = \sum_{t=0}^{T} 2 t \{1-\hat{F}_j(t)\}-\left(\sum_{t=0}^{T} \{1-\hat{F}_j(t)\}\right)^2,
\end{equation*}
and the covariance of the bloom dates between sites $j$ and $j'$ as
\begin{equation*}
    \hat{\bm\Sigma}_{j j^{'}} = \sum_{t_1=0}^{T} \sum_{t_2=0}^{T} \hat{F}_{jj'}(t_1,t_2) - \hat{F}_{j}(t_1) \hat{F}_{j'}(t_2).
\end{equation*}
We then compute the first two eigenvectors of $\hat{\bm\Sigma}$ to get the first two principal components. 

To estimate the covariance matrix, we make a third assumption in addition to the two from Section \ref{subsec:assumptions}. The third assumption (A3) is comonotonicity, $F_{jj'}(t_1,t_2) = F_j(t_1) \wedge F_{j'}(t_2)$. The comonotonicity assumption has the following interpretation. Suppose that (1) one were to hypothetically observe the same plant at two different sites, $j$ and $j'$; (2) that the plant at site $j$ was known to have bloomed before time period $t_1$; and (3) that $F_j(t_1) < F_{j'}(t_2)$. Then the comonotonicity condition states that the flower at site $j'$ must have bloomed before $t_2$. We believe this assumption is reasonable since a flower only blooms after obtaining sufficient resources. Thus, if a flower obtained enough resources to bloom at a site when only a few other flowers have already bloomed, then surely it would have obtained enough resources to bloom at another site when more flowers have already bloomed.

\begin{figure*}[t]
\centering
\includegraphics[width=\linewidth]{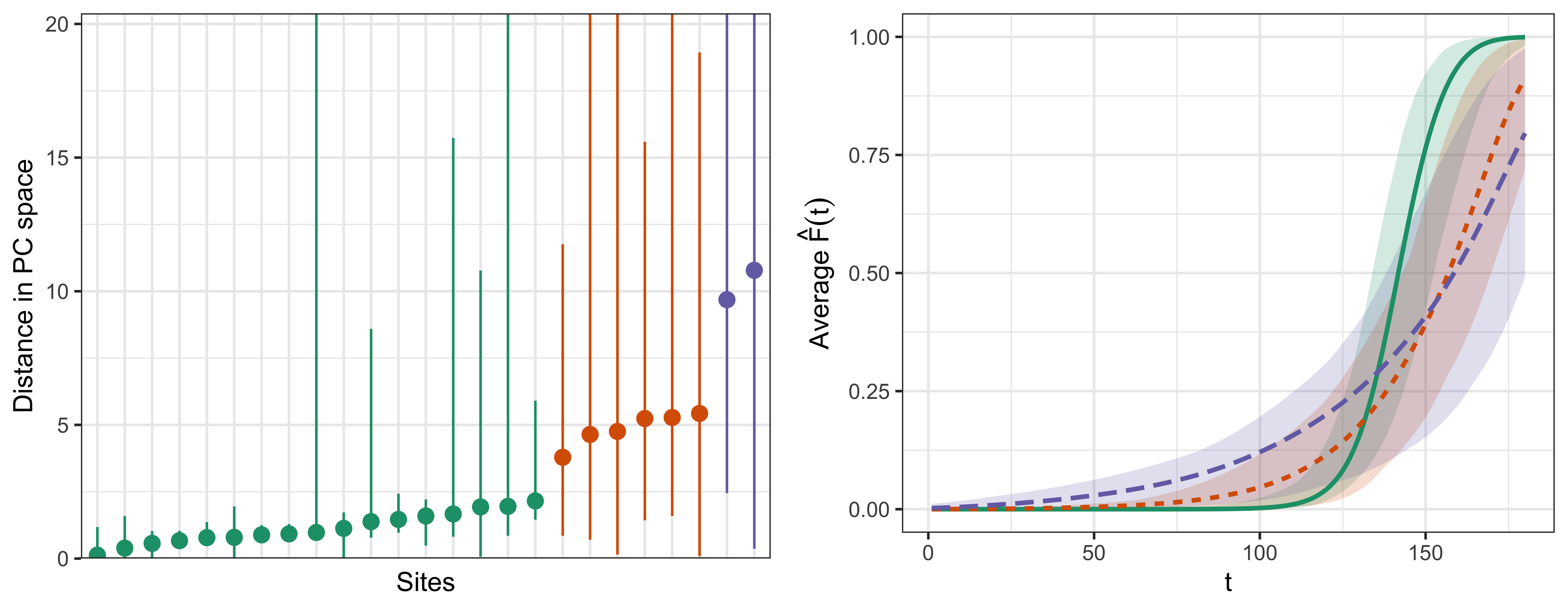}
\caption{Summary of the results from PCA. Left panel: The Mahalanobis distance of each site is calculated from the first two principal components. (Larger distance means more anomalous.) The distance is calculated for each posterior draw from the model, and the distances for the draws that maximized the posterior (dots) and inner 50\% posterior intervals (lines) are displayed. The two sites with the largest PC distance are highlighted in purple. Right panel: The marginal distributions for the two sites with the largest PC distance are displayed (purple) along with the average distribution for the remaining sites (orange and green). The error bands depict the inner 50\% of posterior draws.}
\label{fig:pca-margins-combined}
\end{figure*}

To identify anomalous sites from the covariance matrix, we compute a robust location and scatter estimate of the first two principal components using the FastMCD algorithm \citep{Rousseeuw1999}. For each posterior sample, we record each site's Mahalanobis distance to its center. The results of this PCA analysis are shown in the left panel of Figure~\ref{fig:pca-margins-combined}. The y-axis denotes the distance of each location to the center, and the x-axis denotes the sites, arranged according to the distances given by the sample that maximized the posterior (points). The colors reflect the proximity of each location to the center: close (green), middle (orange), and far (purple). Intervals represent the inner 50\% of posterior draws.

We consider the two purple sites with the largest distances to be anomalously large. The marginal distributions for the anomalous sites are shown in the right panel of Figure~\ref{fig:pca-margins-combined} (purple) along with the average marginal distribution for the remaining sites (orange and green). Although these sites look different, there is considerable uncertainty in how anomalous these sites actually are as illustrated by the large error bounds.  Additional study is necessary to confirm that these sites are in fact anomalous. This could be achieved perhaps by examining additional years or the timing of other seasonal activities at that site.

\section{Discussion}
\label{sec:discussion}

We propose a new method to adjust for the bias that occurs when an individual monitors a fixed location and reports whether an event of interest has occurred or not, such as whether a plant is in flower or not. In Section \ref{sec:background} we provide some background from the field of phenology to motivate the problem; in Section \ref{sec:demonstration}, we show how the timing or frequency of a monitoring pattern produces bias; and in Section \ref{sec:model}, we present a nonparametric Bayesian model that uses monotonic splines to correct for that bias. In Section \ref{sec:application}, we apply the model to data collected by citizen scientists monitoring the flowering status of common lilacs. We find that failing to correct for monitoring bias makes the typical lilac appear to bloom an average of 48 days earlier than it actually did. After correction, two sites had anomalously late bloom dates, although there is considerable uncertainty in how anomalous those sites actually are. Our findings highlight the advantage of using spline smoothing for estimating bloom dates and similar monitoring challenges, as it significantly enhances estimation accuracy and mitigates biases.

Our model fits into a growing statistics literature on citizen science data. This literature has largely focused on data labeling and image annotation \citep{Bonney2014, Nov2014}. Statisticians have made important contributions, such as proposing quality assurance and bias correction methods \citep{Fraisl2022}. Bayesian hierarchical models, in a similar spirit to the method proposed in this paper, have proven valuable in correcting these errors \citep{Santos-Fernandez2021} as has latent factor modeling \citep{Xu2023}. 

Yet citizen science data from regular monitoring, like those considered in this paper, remain largely unexamined by statisticians. While we are motivated by the use of citizen science data for monitoring environments stressed by climate change, similar data arise in a wide variety of settings. For example, many cities use resident complaints to monitor the need for public services such as removing fallen trees or filling potholes. Cities typically rank neighborhoods according to the timing of complaints, but as we demonstrate in this paper, neighborhoods that more promptly report fallen trees or potholes may erroneously appear in greater need of service than neighborhoods that are less prompt.

Citizen science is one of the oldest approaches to data collection, producing some of the longest-running time series \citep{Fraisl2022, Miller-Rushing2012, Kobori2016}. Yet despite their age, citizen science data are experiencing a renaissance --- particularly in the agricultural, biological, and environmental sciences --- due in large part to the increasing sophistication of the technology available to citizen scientists, which allows researchers to quickly employ a vast workforce and quickly collect large quantities of data \citep{Bonney2014, Nov2014, Sherbinin2021}. For example, SciStarter.com hosts thousands of citizen science opportunities. Other examples include the 41 citizen science projects hosted by the National Aeronautics and Space Administration (NASA), eBird \citep{SULLIVAN20092282}, and the Norfolk Bat Survey \citep{NEWSON201538}. 

However, the data collected by citizen scientists cannot automatically be assumed to be representative. Researchers must carefully define the estimand and demonstrate that the estimator based on citizen science data provides a reliable estimate. In many cases, modeling and removing biases, such as monitoring bias, is important because such biases persist when the data are aggregated across space or time, limiting the value of the size and timeliness of the data that can be assembled through the use of citizen scientists.

\bibliographystyle{unsrtnat}
\bibliography{monitoring_bias} 
\end{document}